\documentclass{aa}  

\usepackage{graphicx}
\usepackage{txfonts}

\begin{document}

\title{White dwarfs with hydrogen-deficient atmospheres and the dark matter
       content of the Galaxy}

\author{Santiago Torres\inst{1,2}, 
        Judit Camacho\inst{1},
        Jordi Isern\inst{3,2} \and 
        Enrique  Garc\'{\i}a--Berro\inst{1,2}}
\institute{Departament de F\'\i sica Aplicada, 
           Universitat Polit\`ecnica de Catalunya,
           c/Esteve Terrades 5, 
           08860 Castelldefels, Spain\
           \and       
           Institute for Space  Studies of Catalonia,
           c/Gran Capit\`a 2--4, Edif. Nexus 104,   
           08034  Barcelona,  Spain\ 
           \and
           Institut de Ci\`encies de l'Espai, CSIC,  
           Campus UAB, Facultat de Ci\`encies, Torre C-5, 
           08193 Bellaterra, Spain\ }

\offprints{E. Garc\'\i a--Berro}

\date{\today}

\abstract{The nature  of the  several microlensing events  observed by
          the MACHO  team towards the Large Magellanic  Cloud (LMC) is
          still a subject of  debate.  Low-mass substellar objects and
          stars  with masses larger  than $\sim1\,M_{\sun}$  have been
          ruled  out as  major components  of a  Massive Astrophysical
          Compact Halo  Object (MACHO)  Galactic halo, while  stars of
          half  a  solar mass  seem  to  be  viable candidates.   Main
          sequence stars  have been  already discarded, and  there are
          tight restrictions  on the role played by  white dwarfs with
          hydrogen-dominated atmospheres.}
         {In  this paper  we  evaluate the  contribution  to the  dark
          matter  content   of  the   Galaxy  of  white   dwarfs  with
          hydrogen-deficient atmospheres.}
         {For  this  purpose we  use  a  Monte  Carlo simulator  which
          incorporates  up-to-date  evolutionary  sequences  of  white
          dwarfs    with    hydrogen-rich    and    hydrogen-deficient
          atmospheres. We also take into account detailed descriptions
          of the thick disk and the halo of our Galaxy as well as of a
          reliable model of the LMC.}
         {We  find   that  the  contribution  of   white  dwarfs  with
          hydrogen-deficient  atmospheres   moderately  increases  the
          theoretical estimate  of the  optical depth with  respect to
          the value obtained when  only hydrogen-rich white dwarfs are
          considered.  We also find that the contribuiton of the thick
          disk population  of white dwarfs  is comparable to  the halo
          contribution.  However, the contributions  of both  the halo
          and  the  thick   disk  white-dwarf  populations  are  still
          insufficient to explain the number of events observed by the
          MACHO team.}
         {Finally,  we find  that the  contribution to  the  halo dark
          matter  of the entire  population under  study is  less than
          $10\%$ at the $95\%$ conficence level.}


\titlerunning{White dwarfs with hydrogen-deficient atmospheres and the
              dark matter content of the Galaxy}

\authorrunning{S. Torres et al. }

\maketitle


\section{Introduction}

Since  the pioneering  observational detection  of the  first Galactic
gravitational  microlensing events by  the MACHO  team (Alcock  et al.
1997, 2000), many  efforts have been devoted to  this issue.  In fact,
after these preliminary discoveries,  many other teams pursued similar
studies  to  either confirm  or  discard  their  results. Among  these
observational studies we mention those performed by the EROS (Lasserre
et  al.  2001; Goldman  et al.   2002; Tisserand  et al.   2007), OGLE
(Udalski  et al.   1994), MOA  (Muraki  et al.   1999) and  SuperMACHO
(Becker et  al. 2005)  teams. All of  them have monitored  millions of
stars during  several years in  both the Large Magellanic  Cloud (LMC)
and  the  Small Magellanic  Cloud  (SMC)  to  search for  microlensing
events.  One of the main results of these searches is that none of the
microlensing events found so far has durations between a few hours and
20  days.  This inmediately  translates into  tight contraints  on the
nature of  the objects responsible for the  microlensing events. Today
it is known  that most likely the objects  responsible of the reported
gravitational microlensing  events are stars with  masses ranging from
$\sim 0.1\, M_{\sun}$ and $\sim 1.0\, M_{\sun}$. Thus, for this reason
and because  of their intrinsical  faintness, white dwarfs seem  to be
the best  candidates to explain the observed  microlensing events and,
consequently, they  would also be  obvious candidates to build  up the
baryonic dark matter content of the Galaxy.

In  a series  of previous  papers  we have  exhaustively analyzed  the
contributions  of  the  halo  populations of  carbon-oxygen  (CO)  and
oxygen-neon  (ONe)   white  dwarfs  with   pure  hydrogen  atmospheres
(Garc\'{\i}a--Berro et al.  2004; Camacho et al.  2007).  We have also
extended our  previous studies to  include the population of  halo red
dwarfs (Torres  et al.  2008).   Thus, these studies covered  the full
range of initial masses able to produce microlensing events compatible
with the required durations, and nearly $90\%$ of the stellar content.
The main conclusion  of these papers is that  the entire population of
these stars  can account at most  for $\sim 0.3$ of  the optical depth
found by the  MACHO team.  This in turn  implies that the contribution
of  the  full range  of  masses  between  $0.08$ and  $10\,  M_{\sun}$
represents $\la 5\%$ of the halo  dark matter, with an average mass of
$0.4\, M_{\sun}$.  Even though, we also found that the expected number
of  events obtained  in our  simulations (three  events at  the $95\%$
confidence level) is substantially below the number of events detected
by the  MACHO team.   Thus these results  support the  idea previously
pointed out in several other  studies, that the optical depth found by
the  MACHO   team  is  probably   an  overstimate,  possibly   due  to
contamination of self-lensing objects, variable stars and others.

In all previous  studies in which the contribution  of white dwarfs to
the  dark  matter content  of  the  Galaxy  was analyzed,  white-dwarf
evolutionary sequences with pure hydrogen atmospheres (white dwarfs of
the  DA type)  were employed,  and  the contribution  of non-DA  white
dwarfs was disregarded. However, non-DA white dwarfs represent roughly
20\%  of the  entire  white dwarf  population  and consequently  their
contribution cannot be considered a priori negligible. Moreover, there
is strong  observational evidence that non-DA  white dwarfs respresent
an even  more important  fraction of the  cool white  dwarf population
(Bergeron, Legget  \& Ruiz 2001),  but the current simulations  of the
halo  white dwarf  population  do  not take  this  fact into  account.
Additionally, the colors and magnitudes of cool white dwarfs depend on
their atmospheric  composition --- see  Fig.  1.  Indeed, it  has been
demonstrated  (Hansen  1998)  that  white  dwarfs  with  hydrogen-rich
atmospheres experience a  blue turn at low luminosities,  which is the
result of  extremely strong ${\rm H}_2$  molecular absorption features
in  the  infrared.  This  blue  hook  prevents  DA white  dwarfs  from
reaching very faint magnitudes.  On  the contrary, white dwarfs of the
non-DA types cool  as blackbodies and hence can  reach extremely faint
magnitudes within  the age of  the Galaxy. Again, this  important fact
has been overlooked in the most up-to-date models of the population of
halo white  dwarfs. Finally,  the rate of  cooling of white  dwarfs is
controlled by the thickness and composition of the atmospheric layers.
It  turns  out  that  non-DA  white  dwarfs  cool  faster  than  their
corresponding DA  counterparts, another fact  that has not  been taken
into account in previous simulations.

Another point which deserves attention is whether the lenses belong to
the halo or to an extended thick disk population (Reid, Sahu \& Hawley
2001; Torres et al.  2002).  After  all, and as pointed out by Gyuk \&
Gates  (1999),  the  thick   disk  population  presents  a  reasonable
alternative  to a  halo  population of  lenses. Several  observational
(Oppenheimer  et al.  2001;  Kilic et  al. 2006;  Harris et  al. 2006;
Vidrih et al.   2007) works have addressed this  question, but this is
still a  controversial issue, and  a definitive answer  still requires
more theoretical and observational efforts.

\begin{figure}
\vspace{6.5cm}    
\includegraphics{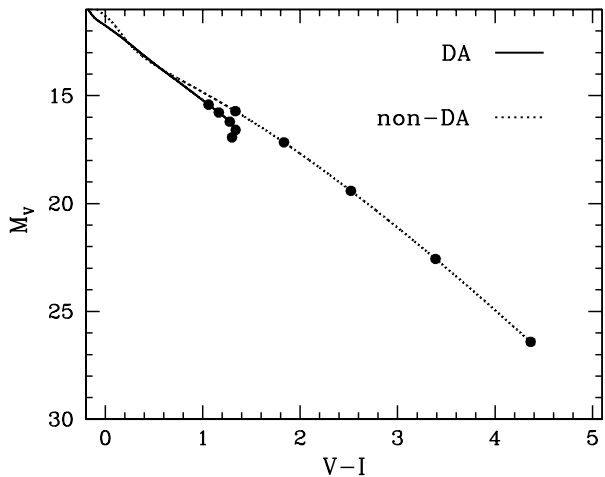} 
\caption{Color-magnitude  diagram of a  typical $0.6\,  M{\sun}$ white
         dwarf. The solid line corresponds  to a cooling sequence of a
         hydrogen-rich (DA)  white dwarf (Salaris et  al. 2000), while
         the dashed  line corresponds to a  helium-rich (non-DA) white
         dwarf (Benvenuto  \& Althaus  1997).  The dots  correspond to
         cooling ages of 6, 8, 10, 12 and 14 Gyr.}
\end{figure}

In this  paper we  extend our previous  studies of the  populations of
halo red and white dwarfs.  In particular, we include the contribution
of  non-DA halo white  dwarfs to  the microlensing  event rate  and we
analyze the role played by  the thick-disk populations of these stars.
Special  emphasis  has been  put  on  the  analysis of  the  simulated
microlensing events as  a function of the range  of colors susceptible
to be  detected by surveys like  EROS and to compare  our results with
those of the MACHO team.  The paper is organized as follows.  In Sect.
2 we summarize the main ingredients  of our Monte Carlo code and other
basic   assumptions   and  procedures   necessary   to  evaluate   the
microlensing optical  depth towards the  LMC. In this section  we also
describe in depth  our thick disk model.  Section 3  is devoted to the
discussion  of our  main results,  including the  contribution  of red
dwarfs  and DA  and non-DA  white dwarfs  to the  microlensing optical
depth towards  the LMC, and to  compare our results with  those of the
EROS team.   In this section we  also estimate the  probability that a
microlensing event  could be assigned to  the thick disk  or the halo,
and  we  discuss the  contribution  of red  and  white  dwarfs to  the
baryonic  content of  the  Galaxy.   Finally, in  Sect.   4 our  major
findings are summarized and we draw our conclusions.


\section{Building the model}

A detailed description  of our Monte Carlo simulator  has been already
presented in Garc\'{\i}a-Berro et al. (1999), Garc\'{\i}a-Berro et al.
(2004), Camacho et al. (2007) and Torres et al. (2002).  Consequently,
we will only summarize the most important inputs here.  At the core of
any Monte  Carlo simulator there  is a pseudo-random  number generator
algorithm, for which we have used  that of James (1990). It belongs to
the  linear  congruential  generator  family.  It  provides  excellent
statistical  properties  and  ensures  a  repetition  period  of  $\ga
10^{18}$, which is  virtually infinite for our purposes.   Each one of
the Monte Carlo simulations presented  here consists of an ensemble of
$\sim5\times10^4$  independent  realizations  of  the  synthetic  star
population, for  which the average  of any observational  quantity and
its corresponding standard deviation were computed.  Here the standard
deviation  means the  ensemble mean  of the  sample dispersions  for a
typical sample.

The main sequence  mass is obtained by drawing  a pseudo-random number
according to  an initial mass  function. We have adopted  the standard
initial mass function of  Scalo (1998).  Other biased non-conventional
initial mass functions (Adams \& Laughlin 1996; Chabrier et al.  1996)
have been disregarded since these mass functions are incompatible with
the observed properties  of the halo white dwarf  population (Isern et
al.  1998; Garc\'{\i}a--Berro et  al.  2004), with the contribution of
thermonuclear  supernovae  to the  metallicity  of  the Galactic  halo
(Canal et al.   1997), and with the observations  of galactic halos in
deep galaxy surveys (Charlot \& Silk  1995). Once the mass of the star
is  chosen, its  main-squence lifetime  is derived  (Iben  \& Laughlin
1989),  and we are  able to  determine which  stars have  evolved into
white  dwarfs  or  remain  in  the main-squence  as  red  dwarfs.   We
considered   red   dwarfs   to   have   masses   in   the   range   of
$0.08<M/M_{\sun}<1$.  For these stars we have adopted the evolutionary
models of Baraffe  et al.  (1998).  Stars with  such small masses have
long  main-sequence lifetimes  and,  therefore, no  post-main-sequence
evolutionary tracks were required. For those stars which have had time
enough to enter into the white  dwarf cooling track and given a set of
theoretical  cooling   sequences  and   the  initial  to   final  mass
relationship  (Iben \&  Laughlin 1989)  their  luminosities, effective
temperatures and  colors were obtained. The  cooling sequences adopted
here depend on the mass of  the white dwarf.  White dwarfs with masses
smaller than $M_{\rm WD}=1.1\, M_{\sun}$ are expected to have CO cores
and, consequently, we adopt for  them the cooling tracks of Salaris et
al.   (2000) if  they belong  to  the DA  spectral class.  If, on  the
contrary, the  white dwarf has a hydrogen-deficient  atmosphere we use
the cooling sequences of Benvenuto \& Althaus (1997) --- corresponding
to  pure helium  atmospheres  --- and  the  bolometric corrections  of
Bergeron,  Wesemael  \&  Beauchamp  (1995).  These  are  our  fiducial
cooling sequences.  However, to study the effects of different cooling
sequences for  hydrogen-deficient white dwarfs  (and, hence, different
cooling speeds) which  may affect our results we  also use the cooling
sequences of Bergeron et al.  (1995) --- see below.  White dwarfs with
masses larger than $M_{\rm  WD}=1.1\, M_{\sun}$ most probably have ONe
cores, and  for these white dwarfs  we adopt the  cooling sequences of
Althaus et  al.  (2007).  All these cooling  sequences incorporate the
most  accurate physical  inputs  for the  stellar interior  (including
neutrinos, crystallization,  phase separation and  Debye cooling) and,
for  the case  of DA  white  dwarfs, reproduce  the blue  turn at  low
luminosities (Hansen 1998).

\subsection{The fraction of DA and non-DA white dwarfs}

To assign a spectral type to each of the white dwarfs in the simulated
sample  we proceeded as  follows.  In  a first  set of  simulations we
adopted the canonical fraction of 80\% of white dwarfs of the spectral
type DA and  20\% of the non-DA class,  independently of the effective
temperature of the white dwarf.  We regard this as our fiducial model,
and we refer to it  as model A. However, several observations indicate
that  this ratio  is a  function  of the  effective temperature.   For
instance,  the well-known  DB-gap, where  no  white dwarfs  of the  DB
spectral class can be  found, occurs at effective temperatures between
45\,000  K and  30\,000 K.   Additionally, Bergeron,  Leggett  \& Ruiz
(2001) found  that  most  white  dwarfs  with  effective  temperatures
ranging  from 6\,000  K to  5\,000 K  are DAs.   Finally,  Bergeron \&
Leggett (2002) argued that all  white dwarfs cooler than 4\,000 K have
mixed  H/He  atmospheres.  Many  of  these  early  findings have  been
corroborated by the wealth of data obtained from recent large surveys,
like the Sloan  Digital Sky-Survey (Harris et al.   2006; Kilic et al.
2006).   Accordingly,   we  have  have   produced  a  second   set  of
simulations,  and  we  refer  to  them as  model  B,  following  these
observational  results.   Basically, in  model  B  we  adopt the  same
fraction  of DA  white  dwarfs (80\%)  for  temperatures above  6\,000
K. All  white dwarfs  in the range  of effective  temperatures between
6\,000 K and 5\,000 K were considered to be DA white dwarfs.  Finally,
for effective  temperatures below  this value we  adopt a  fraction of
50\% (Bergeron \& Legget 2002; Gates  et al.  2004).  We would like to
note  that we  model the  transitions between  the  different spectral
classes  in a  purely heuristic  way  because currently  there are  no
cooling sequences which correctly reproduce these transitions, as this
is a  long-standing problem, which is  indicative of a  failure of the
theoretical cooling  models.  However, our  model correctly reproduces
the  observations, and  thus we  consider it  to be  a  fair approach.
Finally,  to check  the  sensitivity  of our  results  to the  adopted
cooling tracks we have also computed a third set of simulations, based
on  model B,  in which  we use  the cooling  sequences of  Bergeron et
al. (1995). We refer to this model as model C.

\subsection{The halo model}

We have adopted a spherically symmetric halo.  In particular the model
used here  is the typical  isothermal sphere of  a radius of  $5$ kpc,
also called  the ``S-model'', which  has been extensively used  by the
MACHO collaboration  (Alcock et al.   2000; Griest 1991).  Despite the
fact  that other  models  as for  instance  the exponential  power-law
models or  the Navarro, French  \& White (1997) density  profiles have
been  proposed, our  studies  (Garc\'{\i}a--Berro et  al.  2004)  have
shown that  no relevant  differences are found  when these  models are
used.   Furthermore, we  do not  consider non-standard  models  of the
Galactic halo, such as  models with flattened density profiles, oblate
halo models  and others  because a thorough  study of these  models is
beyond the scope of this paper.

The kinematical  properties of the  halo population have  been modeled
according to Gaussian  laws (Binney \& Tremaine 1987)  with radial and
tangential  velocity  dispersions  accordingly  related by  the  Jeans
equation and  fulfilling the  flat rotation curve  of our  Galaxy.  We
have  adopted  standard  values  for  the  circular  velocity  $V_{\rm
c}=220$~km/s  as  well  as  for  the  peculiar  velocity  of  the  Sun
$(U_{\sun},  V_{\sun},W_{\sun})=(10.0,  15.0,  8.0)$~km/s  (Dehnen  \&
Binney 1998).  Besides, we  have rejected stars with velocities higher
than 750~km/s, because they  would have velocities exceeding 1.5 times
the escape velocity.  Finally, since  white dwarfs usually do not have
determinations  of the radial  component of  the velocity,  the radial
velocity is  eliminated when a comparison with  the observational data
is needed.

Finally, to compare the simulated results with the observational ones,
a normalization criterion should be used.  We have proceeded as in our
previous  papers  (Camacho et  al.   2007;  Garc\'{\i}a--Berro et  al.
2004; Torres et al. 2008). That is, we have normalized our simulations
to the local density of halo white dwarfs obtained from the halo white
dwarf luminosity  function of  Torres et al.   (1998), but  taken into
account the new  halo white dwarf candidates found  in the SDSS Stripe
82  (Vidrih  et al.   2007).   Nevertheless,  we  emphasize that  when
normalizing to the  local density of halo white  dwarfs obtained using
the white dwarf luminosity function  we only consider those stars with
velocities higher  than $250$~km/s, given that only  those stars would
be genuinely considered as halo members and would be used to build the
observational halo  luminosity function (Liebert et  al.  1989; Torres
et  al.  1998).  This  is totally  equivalent  to the  adopted cut  in
reduced proper motion employed by Flynn et al.  (2001).  Additionally,
only the number density of DA white dwarfs was considered to normalize
the simulations, since all but one  of the white dwarfs used to obtain
the  luminosity function  of  Torres et  al.   (1998) were  of the  DA
spectral  type. Obviously, imposing  this normalization  we implicitly
assume that the MACHO results and the direct surveys are complementary
and seem  to be probing the  same populations, whatever  the nature of
those populations (Hansen \& Liebert 2003).

\subsection{The thick disk model}

The structure and  kinematics of the Galactic disk  remain a source of
controversy and discussion. In particular the nature of the thick disk
is  an  active field  of  research.   Consequently  we have  used  two
different models in our simulations. The first of these is a canonical
thick  disk   model,  which  we  consider  as   a  starting  reference
model. However, there  are alternative thick disk models  based on the
kinematics of  metal-poor stars of the  Galaxy --- see  Chiba \& Beers
(2000) and references therein ---  that challenge the canonical model.
Accordingly,  we also  consider the  most recent  thick disk  model of
Carollo et al.  (2009), which is  based on the SDSS Data Release 7. We
describe them separately.

The kinematical properties of the canonical model are well represented
by an ellipsoid  with constant values of the  velocity and dispersions
and an asymmetrical drift.  Within this model the spatial distribution
is  generally assumed to  follow exponential  laws characterized  by a
scale height and  a scale length with no  vertical gradients.  Thus we
have choosen a double exponential  law for the density profile of this
model with  a scale height of  1.5 kpc and  a scale length of  3.0 kpc
(Reid 2005).  The kinematical properties  of the synthetic  thick disk
stars have  been modeled according  to an ellipsoid with  the standard
dispersion $(\sigma_U, \sigma_V, \sigma_W)=(60, 45, 35) \; {\rm km} \;
{\rm s}^{-1}$ and an asymmetric velocity drift $V_\phi=-40\;{\rm km}\;
{\rm s}^{-1}$ (Reid 2005).

In a recent  study, Carollo et al.  (2009)  analyzed the structure and
kinematical properties of the Milky Way based on the Sloan Digital Sky
Survey Data Release 7 and  showed evidence that a sizeable fraction of
the  thick  disk is  composed  by  metal-weak  stars with  independent
kinematical properties.   Following Carollo et  al.  (2009) we  use an
ellipsoid    with   standard    dispersions    $(\sigma_U,   \sigma_V,
\sigma_W)=(59, 40, 44)\; {\rm km}\;  {\rm s}^{-1}$, and a scale height
and  a scale  length  of 1.36  and  2.0 kpc,  respectively.  The  most
distinctive feature of the model of  Carollo et al. (2009) is that the
asymmetric drift  varies as  a function of  height above  the Galactic
plane.   Specifically,  the  gradient   in  the  asymmetric  drift  is
$\Delta\langle   V_{\phi}\rangle/\Delta|z|=-36\;  {\rm  km}   \;  {\rm
s}^{-1} \; {\rm kpc}^{-1}$, which  agrees with the previous studies of
Chiba  \&  Beers  (2000).   We  note that  although  the  observations
indicate that only a fraction of  the thick disk could be explained by
the  metal-weak  thick disk  population,  we  have  considered a  full
metal-weak  thick  disk to  obtain  an  upper  limit to  the  possible
contribution of this population to the microlensing experiments.

Additionally, in both cases we took into account the peculiar velocity
of the Sun $(U_{\sun}, V_{\sun},W_{\sun})=(10.0, 15.0, 8.0)\;{\rm km\;
s^{-1}}$ (Dehnen \& Binney 1998) and discarded those stars that escape
the  potential of the  Galaxy.  We  also assumed  that the  thick disk
formation  started 12  Gyr  ago  with a  maximum  star formation  rate
occuring 10  Gyr ago and exponentially decreased  since, following the
model of Gilmore, Wyse \& Jones (1995).  Finally our thick disk models
have been  normalized assuming that the thick  disk density represents
$8.5\%$ of the thin disk density (Reid 2005).

\subsection{The LMC model}

In order to mimic the microlensing experiments towards the LMC we have
simulated it  following closely the detailed LMC  descriptions of Gyuk
et al.  (2000)  and Kallivayalil et al. (2006).   Our model takes into
account among  other parameters the  scale length and scale  height of
the LMC,  its inclination and its kinematical  properties.  This model
provides us with a synthetic population of stars representative of the
monitored point  sources.  Afterwards we  evaluate which stars  of the
Galactic halo could  be responsible of a microlensing  event.  We have
only considered stars fulfilling a  series of conditions. First of all
the lensing  stars should be  fainter than a certain  magnitude limit.
In a  second step we have checked  if the lens is  inside the Einstein
tube  of  the monitored  star.  That is,  we  checked  if the  angular
distance between the  lens and the monitored star  is smaller than the
Einstein radius. We recall here that the Einstein radius is given by

\begin{equation}
R_{\rm E}=2\sqrt{\frac{GMD_{\rm OS}}{c^2}x(1-x)}
\end{equation}

\noindent where $D_{\rm OS}$ is the observer-source distance, $x\equiv
D_{\rm OL}/D_{\rm OS}$ and $D_{\rm OL}$ is the observer-lens distance.
Finally, we  filtered those  stars which are  candidates to  produce a
microlensing   event   with   the   detection   efficiency   function,
$\varepsilon(\hat  t_i)$,  where  $\hat  t_i$  is  the  Einstein  ring
diameter  crossing  time.  The  detection  efficiency  depends on  the
particular  characteristics of the  experiment.  In  our case  we have
reproduced the MACHO and EROS experiments.  Specifically for the MACHO
collaboration we have  taken $1.1\times 10^7$ stars during  5.7 yr and
over $13.4\ {\rm
 deg^2}$,  whereas the detection efficiency has been
modeled according to:

\begin{figure*}
\vspace{11.5cm}    
\includegraphics{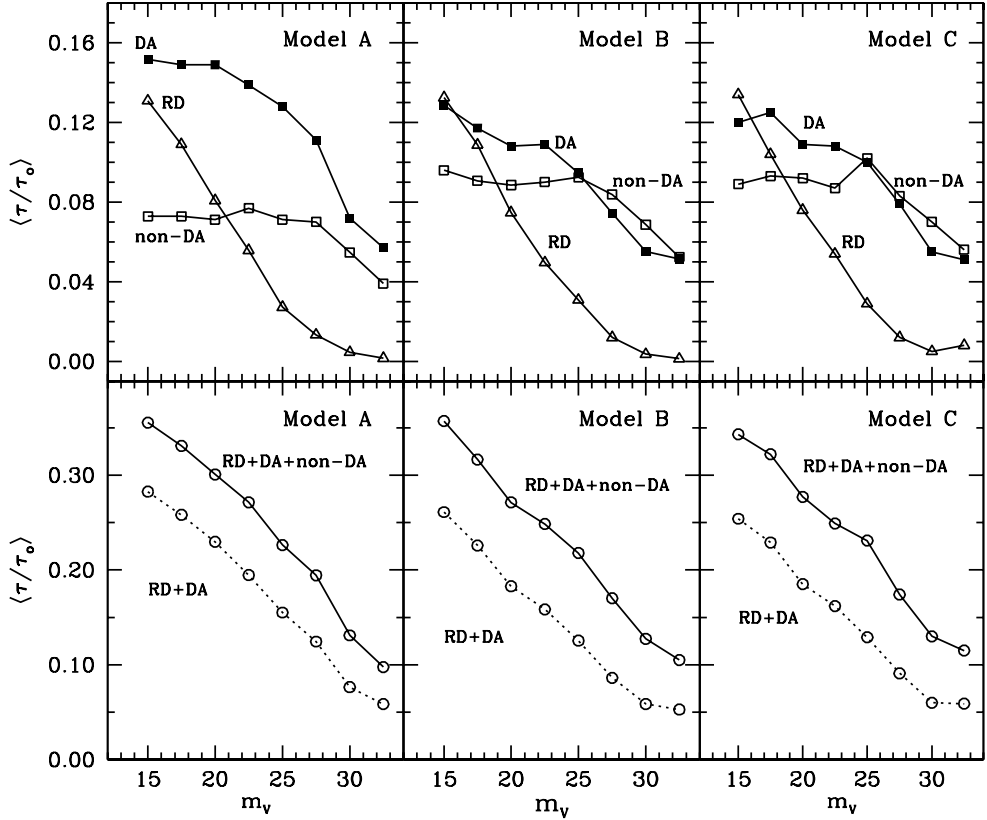} 
\caption{Microlensing optical  depth towards the LMC as  a function of
         the limiting magnitude. Solid  and open squares represent the
         DA  and non-DA  white dwarf  populations,  respectively.  Red
         dwarfs are  represented by  open triangles, while  the entire
         population is shown by open circles.}
\end{figure*}

\begin{equation}
\varepsilon(\hat{t})= 
\left\{
\begin{array}{cc}
0.43\,{\rm e}^{-(\ln(\hat{t}/T_{\rm m}))^{3.58}/0.87}, &\hat{t}>T_{\rm m} \\   
43\,{\rm e}^{-|\ln(\hat{t}/T_{\rm m})|^{2.34}/11.16},  &\hat{t}<T_{\rm m},
\end{array}
\right.
\end{equation}

\noindent where $T_{\rm m}=250$ days.  This expression provides a good
fit to the results of Alcock  et al.  (2000).  For the EROS experiment
we  have  used $0.7\times  10^7$  stars over  a  wider  field of  $84\
{\deg^2}$  and over  a period  of $6.7$  yr.  Regarding  the detection
efficiency we have adopted a numerical fit to the results of Tisserand
et al. (2007).

For all  the simulations presented  here, we extracted  the parameters
relevant  to characterize  the microlensing  experiments.   A complete
description  of the  various  parameters which  are  of importance  in
discussing  gravitational microlensing  can be  found in  Mollerach \&
Roulet (2002)  and Schneider et al.  (2004).   Among these parameters,
perhaps the most important one  for our purposes is the optical depth,
$\tau$, which measures the probability of  a star to be magnified by a
lens  at  a given  time  by  more than  a  factor  of  1.34.  From  an
observational  point of  view an  estimate  of this  parameter can  be
obtained by the expression (Alcock et al. 2000):

\begin{equation}
\tau=\frac{1}{E}\frac{\pi}{4}\sum_i 
\frac{\hat t_i}{\varepsilon(\hat t_i)},
\end{equation}

\noindent where $E$  is the total exposure in  star-years. The optical
depth  is  independent  of  the  lens motion  and  mass  distribution.
However, since the experiments measure  the number of events and their
durations,   additional  information   can  be   obtained   using  the
microlensing rate $\Gamma$  and its distribution as a  function of the
event durations.  This parameter  represents nothing else but the flux
of lenses  inside the microlensing  tube. Finally, an estimate  of the
expected number of events can be achieved by the expression

\begin{equation}
N_{\rm exp}=E\int_{0}^{\infty}\frac{d\Gamma}{d\hat{t}}\varepsilon(\hat
t_i)d\hat t_i.
\end{equation}


\section{Results}

\subsection{The optical depth towards the LMC}

\begin{table*}
\centering
\begin{tabular}{lrrrrrrrrrrrrr}
\hline
\hline
\multicolumn{1}{c}
{\ } 
&\multicolumn{4}{c}{Model A} 
&\multicolumn{4}{c}{Model B} 
&\multicolumn{4}{c}{Model C}\\
 \hline 
Magnitude & 17.5 & 22.5 & 27.5 &32.5 & 17.5  & 22.5 & 27.5 & 32.5 & 17.5  & 22.5 & 27.5 & 32.5 \\
\cline{2-5}
\cline{6-9}
\cline{10-13}
$\langle N_{\rm WD}\rangle$                       & $0\pm1$ & $0\pm1$ & $0\pm1$ & $0\pm1$ & $0\pm1$ & $0\pm1$ & $0\pm1$ & $0\pm1$ & $0\pm1$ & $0\pm1$ & $0\pm1$ & $0\pm1$\\
$\langle N_{\rm RD}\rangle$                       & $0\pm1$ & $0\pm1$ & $0\pm1$ & $0\pm1$ & $0\pm1$ & $0\pm1$ & $0\pm1$ & $0\pm1$ & $0\pm1$ & $0\pm1$ & $0\pm1$ & $0\pm1$\\
$\langle m_{\rm WD}\rangle$ $(M/M_{\sun})$        & 0.599   & 0.605   & 0.627   & 0.721   & 0.604   & 0.602   & 0.627   & 0.661   & 0.600   & 0.605   & 0.619   & 0.678  \\
$\langle m_{\rm RD}\rangle$ $(M/M_{\sun})$        & 0.325   & 0.233   & 0.109   & 0.081   & 0.319   & 0.227   & 0.118   & 0.080   & 0.315   & 0.227   & 0.124   & 0.079  \\
$\langle\eta\rangle$                              & 0.204   & 0.227   & 0.323   & 0.837   & 0.810   & 0.834   & 0.927   & 0.993   & 0.780   & 0.828   & 0.913   & 0.987  \\
$\langle\mu\rangle\;(''\,{\rm yr}^{-1})$          & 0.020   & 0.014   & 0.010   & 0.009   & 0.021   & 0.016   & 0.015   & 0.014   & 0.021   & 0.014   & 0.012   & 0.016  \\
$\langle d\rangle$ (kpc)                          & 2.54    & 3.62    & 5.28    & 5.26    & 2.29    & 3.28    & 3.49    & 3.88    & 2.37    & 3.71    & 4.09    & 3.02   \\
$\langle V_{\rm tan} \rangle\;({\rm km\,s}^{-1})$ & 243     & 247     & 253     & 244     & 239     & 244     & 252     & 260     & 239     & 252     & 241     & 238   \\
$\langle\hat{t}_{\rm E}\rangle $ (d)              & 41.3    & 49.6    & 59.0    & 60.7    & 39.1    & 45.3    & 47.7    & 60.9    & 40.1    & 45.6    & 52.1    & 65.5  \\
$\langle\tau/\tau_0\rangle$                       & 0.331   & 0.271   & 0.194   & 0.098   & 0.316   & 0.257   & 0.179   & 0.109   & 0.321   & 0.266   & 0.174   & 0.115 \\
\hline
\hline 
\end{tabular}
\caption{Summary of the results  obtained for the entire population of
         microlenses towards  the LMC for several  magnitude cuts when
         the results of the MACHO collaboration are simulated.  An age
         of the halo of 14~Gyr has been adopted.}
\end{table*}

The optical  depth provides the most immediate  and simple information
about the  microlensing experiments.  Thus we compare  our simulations
with the optical depth derived  by the MACHO collaboration.  In Fig. 2
we  show  the contribution  to  the  optical  depth of  the  different
populations under study as a function of the adopted magnitude cut, in
the same  manner as it was  done in Garc\'{\i}a--Berro  et al.  (2004)
and subsequent  papers.  Our simulations  have been normalized  to the
value  derived by  Alcock et  al. (2000),  $\tau_0=1.2\times 10^{-7}$.
The contributions  to the microlensing optical depth  of the different
populations  are  represented  by  solid  and  open  squares  for  the
populations  of DA  and non-DA  white dwarfs  respectively,  while the
contribution  of  the  red  dwarf  population  is  displayed  by  open
triangles.   Finally,  the contribution  to  the microlensing  optical
depth of the entire population is  shown by open circles. For the sake
of clarity, the contribution of  the ONe white dwarf population is not
shown in the top  panels of Fig. 2, but it is  taken into account when
the  total contribution (shown  in the  bottom panels  of Fig.   2) is
computed.  As  can be seen, for  model A at bright  magnitude cuts the
contribution to the microlensing optical  depth is roughly 7\% for the
population of  non-DA white  dwarfs and $\sim  15\%$ for the  DA white
dwarf population (see the top left  panel of Fig.  2).  However, it is
remarkable that  as the magnitude  cut increases, the  contribution of
non-DA white dwarfs remains almost constant, while the contribution of
DA white  dwarfs rapidly drops.  This  is a direct  consequence of the
faster cooling rate of non-DA white dwarfs and of the fact that non-DA
white dwarfs do not experience the blue turn.  The contribution of the
red dwarf population is very similar  to the one found in our previous
studies, with a fairly constant  decreasing slope as the magnitude cut
increases.   The   decreasing  contribution  of  red   dwarfs  to  the
microlensing optical  depth for  increasing magnitude cuts  stems from
the fact  that in general red  dwarfs are brighter  than regular white
dwarfs.  When  model B is  considered, the overall contribution  of DA
white dwarfs is  smaller than the contribution of  non-DA white dwarfs
--- see the top  central panel of Fig.  2.  Note that  for model A the
opposite occurs, that is, the  contribution of DA white dwarfs is more
sizeable than the one of non-DA  white dwarfs.  The reason for this is
easy  to understand.   Since  the luminosity  function  of halo  white
dwarfs of Torres et al. (1998) only provides the density of relatively
bright DA  white dwarfs ($\log(L/L_{\sun})\ga -3.7$)  and the fraction
of low  luminosity white dwarfs in  model B is only  50\% (in contrast
with  that of  model  A, for  which  a fraction  of  80\% was  adopted
independently of  the effective temperature), the  contribution of low
luminosity white dwarfs to  the optical depth decreases.  Finally, the
top  right panel of  Fig. 2  shows the  relative contributions  to the
microlensing  optical depth  when model  C is  considered.  As  can be
seen, the  results are  virtually indistinguishable of  those obtained
for model  B, as one  should expect given  that the cooling  tracks of
Bergeron  et al.  (1995)  are very  similar to  those of  Benvenuto \&
Althaus (1997).  In  all cases it is important to  realize that as far
as the entire population is  concerned, there is a noticeable increase
in the contribution to the  optical depth, which is exclusively due to
the inclusion  of non-DA white  dwarfs in our  calculations. Moreover,
the global  contribution of white  dwarfs to the  microlensing optical
depth is  very similar  in both  models --- see  the bottom  panels of
Fig.  2 ---  for magnitude  cuts larger  than $m_V\sim  23^{\rm mag,}$
which is a reasonable value  for current surveys.  This value, roughly
$30\%$  of the  observed optical  depth  obtained by  the MACHO  team,
represents a $50\%$  increment with respect to the  value found in our
previous studies, see Torres et al. (2008).

A more  detailed information can  be obtained from our  simulations. A
summary is  presented in  Table 1, where  we show  several interesting
parameters  for the  three models  under study  as a  function  of the
adopted  magnitude cut.   In  particular  we show  in  this table  the
expected  number of  white dwarf  microlensing events,  the  number of
microlensing events  produced by red  dwarfs, the average mass  of the
microlenses for both the microlensing events produced by white and red
dwarfs, the  fraction of the white dwarf  microlensing events produced
by white  dwarfs of the non-DA  spectral type ($\eta$)  over the total
white dwarf  microlensing events, the average  proper motion, distance
and  tangential velocity  of  the lenses,  the corresponding  Einstein
crossing   times  of   the  microlenses   and  finally   the  relative
contribution to the microlensing  optical depth. A close inspection of
Table 1 reveals  that all three models produce  similar results except
in one aspect,  the fraction of microlensing events  attributable to a
non-DA  white  dwarf.   As  can   be  seen,  the  expected  number  of
microlensing events is very small in all models, since in all cases no
more  than one  microlensing  event is  expected  to be  found at  the
$1\sigma$ confidence  level. Additionally,  the average masses  of the
microlenses are around  $0.6\, M_{\sun}$ in the case  of white dwarfs,
while  for  red  dwarfs  it  is $\sim  0.2\,  M_{\sun}$,  the  average
distances to  the microlenses are  also very similar for  both models,
and  there are  no significant  differences in  the  Einstein crossing
times.  The  only relevant difference  between the simulations  is the
spectral  type  of  the  white  dwarf responsible  for  the  simulated
microlensing  events.  Whereas  for model  A the  DA type  prevails in
$\sim 73\%$  of the cases,  for model B  this fraction drops  to $\sim
20\%$ of the cases, while for  model C we obtain a very similar value,
$\sim 21\%$.   This can be  understood by the same  reasoning employed
before.   For  models B  and  C,  at  low effective  temperatures  the
fraction of  hydrogen-rich white  dwarfs is considerably  smaller than
for model A  and, additionally, old DA white  dwarfs are brighter than
non-DAs.  Thus for models B and  C non-DA white dwarfs dominate at low
luminosities and produce most of the microlensing events.

Additionally, from a detailed analysis of the data used to build Table
1, we  have found that on  average the microlenses  produced by non-DA
white  dwarfs  have  slightly  higher  average  masses  ($\sim  0.61\,
M_{\sun}$ and  $\sim 0.56\, M_{\sun}$, respectively) and  can be found
at  a  smaller   distances  ($\sim  1.7$  kpc  and   $\sim  2.9$  kpc,
respectively)  than  those produced  by  the  population  of DA  white
dwarfs.  That  is again a  consequence of the different  cooling rates
and colors  of non-DA white  dwarfs.  As previously  mentioned, non-DA
white  dwarfs cool  faster and  moreover,  as they  cool, they  become
substantially dimmer than their corresponding DA counterparts.  Hence,
the  population of  non-DA  white dwarfs  can  produce microlenses  at
significantly  smaller distances.   Also, these  values do  not depend
significantly  on   the  model  adopted  for  the   evolution  of  the
atmospheric composition  of white  dwarfs.  Since the  distribution of
velocities does not  depend on the spectral type,  the final result is
that  the Einstein  crossing times  are on  average different  for the
microlensing events produced by non-DA  and DA white dwarfs ($\sim 40$
and $\sim 57$ days, respectively).

\begin{figure}
\vspace{14.5cm}
\includegraphics{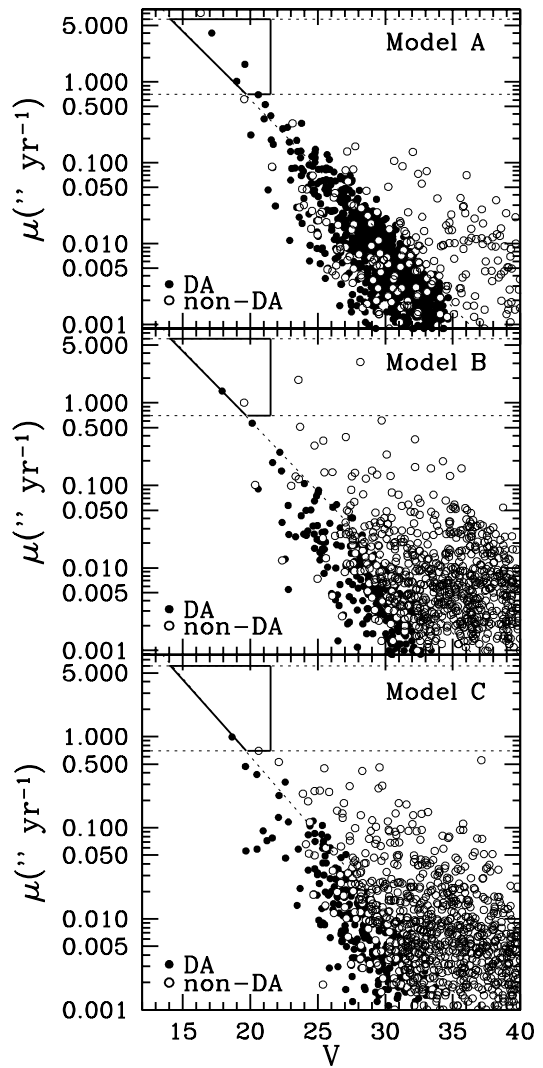}
\caption{Proper motion as a function of the $V$ magnitude for the halo
         white dwarf population. The EROS selection zone is shown as a
         bold solid line.}
\end{figure}

\begin{table*}[ht]
\centering
\begin{tabular}{lrrrrrrrrrrrrr}
\hline
\hline
\multicolumn{1}{c}
{\ } 
&\multicolumn{4}{c}{Model A} 
&\multicolumn{4}{c}{Model B} 
&\multicolumn{4}{c}{Model C}\\
\hline 
Magnitude                                          & 17.5    & 22.5    & 27.5    & 32.5    & 17.5    & 22.5    & 27.5    & 32.5    & 17.5    & 22.5    & 27.5    & 32.5 \\
\cline{2-5} 
\cline{6-9} 
\cline{10-13} \\
$\langle N_{\rm WD}\rangle$                        & $0\pm1$ & $0\pm1$ & $0\pm1$ & $0\pm1$ & $0\pm1$ & $0\pm1$ & $0\pm1$ & $0\pm1$ & $0\pm1$ & $0\pm1$ & $0\pm1$ & $0\pm1$\\
$\langle N_{\rm RD}\rangle$                        & $0\pm1$ & $0\pm1$ & $0\pm1$ & $0\pm1$ & $0\pm1$ & $0\pm1$ & $0\pm1$ & $0\pm1$ & $0\pm1$ & $0\pm1$ & $0\pm1$ & $0\pm1$\\
$\langle m_{\rm WD}\rangle$ $(M/M_{\sun})$         & 0.383   & 0.601   & 0.618   & 0.744   & 0.602   & 0.598   & 0.621   & 0.674   & 0.590   & 0.605   & 0.610   & 0.666  \\
$\langle m_{\rm RD}\rangle$ $(M/M_{\sun})$         & 0.328   & 0.206   & 0.111   & 0.083   & 0.305   & 0.208   & 0.118   & 0.082   & 0.304   & 0.177   & 0.112   & 0.085  \\
$\langle\eta\rangle$                               & 0.186   & 0.246   & 0.373   & 0.671   & 0.774   & 0.800   & 0.919   & 0.965   & 0.711   & 0.882   & 0.957   & 0.976  \\
$\langle \mu\rangle\;(''\,{\rm yr}^{-1})$          & 0.021   & 0.017   & 0.009   & 0.006   & 0.022   & 0.016   & 0.016   & 0.015   & 0.021   & 0.017   & 0.011   & 0.016  \\
$\langle d\rangle$ (kpc)                           & 2.51    & 3.16    & 5.52    & 9.04    & 2.46    & 3.35    & 3.58    & 3.80    & 2.56    & 3.23    & 4.83    & 3.21   \\
$\langle V_{\rm tan} \rangle\; ({\rm km\,s}^{-1})$ & 250     & 258     & 244     & 245     & 254     & 259     & 272     & 275     & 261     & 257     & 259     & 245    \\
$\langle\hat{t}_{\rm E}\rangle $ (d)               & 41.9    & 44.1    & 60.7    & 88.8    & 35.9    & 40.9    & 43.5    & 53.2    & 38.1    & 39.6    & 49.4    & 63.9   \\
$\langle\tau/\tau_0\rangle$                        & 0.977   & 0.810   & 0.659   & 0.384   & 0.794   & 0.678   & 0.400  & 0.214    & 0.775   & 0.618   & 0.352   & 0.259  \\
\hline
\hline 
\end{tabular}
\caption{Summary of the results  obtained for the entire population of
         microlenses towards  the LMC  for the EROS  experiment, using
         models A, B  and C and adopting an age of  the halo of 14~Gyr
         and several magnitude cuts.}
\end{table*}

\subsection{The EROS experiment}

The EROS experiment has monitored  a wider solid angle and less crowed
fields in LMC than the MACHO team.  In addition, it has also monitored
the SMC.   For these reasons, self-lensing  of the LMC  should be less
important  in  the EROS  experiment  than in  the  case  of the  MACHO
collaboration.   Consequently, a  smaller value  of the  optical depth
should be  expected, and this is  indeed the case.   The EROS results,
adopting   a    standard   halo   model    and   assuming   $\tau_{\rm
SMC}=1.4\tau_{\rm LMC}$  indicate that the  microlensing optical depth
is $\tau_0=0.36\times10^{-7}$ (Tisserand et  al.  2007), which is four
times smaller than that obtainded by the MACHO team.

We have performed a set of simulations emulating the conditions of the
EROS    experiment    using    the    same    populations    described
previously. Although  only small differences should  be expected, this
new series of  simulations represents a test of  the robustness of our
numerical procedures.   In Table 2  we summarize the  results obtained
for this set of  simulations.  Our simulations show that independently
of the adopted model for the  spectral type of white dwarfs, the joint
population  of  red dwarfs  and  white  dwarfs  of the  galactic  halo
provides at  most $\sim  90\%$ of the  optical depth estimated  by the
EROS  team. This  value represents  an  increase of  $\sim 20\%$  with
respect to  the one  obtained in our  previous simulations  (Torres et
al. 2008). Obviously, the non-DA white dwarf population is responsible
for this result, and this  confirms our previous conclusion that there
is a general agreement between  the theoretical models and the results
of the EROS team.

Moreover the EROS  experiment used a set of  selection criteria in the
search of  halo white  dwarfs to distinguish  halo objects  from thick
disk stars (Goldman et al.  2002). For those stars detectable by EROS,
namely those with magnitudes  brighter than $V=21.5$ and $I=20.5$, the
selection criteria are implemented by two cuts. The first one uses the
reduced proper motion and requires that the reduced proper motion of a
halo object  should be $H_V>22.5$.  The  second cut is  applied to the
resulting  sample  and only  selects  those  stars  with large  proper
motions, $\mu>0.8''\,{\rm  yr^{-1}}$. In Fig.  3 we  present a typical
simulation  of  the  halo   white  dwarf  population  adapted  to  the
requirements of the EROS  team.  The previously mentioned criteria are
displayed by dotted lines, while  the resulting halo selection zone is
represented by a  bold solid line.  For model A we  obtain that at the
$1\sigma$ confidence level, $4\pm2$ DA white dwarfs and $1\pm1$ non-DA
white dwarfs should be found in  the selection zone, while for model B
we obtain $1\pm1$ and $1\pm1$ white dwarfs, respectively, and the same
occurs for model C.  These results indicate that the models which take
into account  the temperature dependence  of the white  dwarf spectral
type (models  B and C) seem  to yield a more  realistic and consistent
estimate, given that it agrees well  with the null results of the EROS
team.  It is also worth noting that the applied selection criteria, in
particular the proper motion cut ($\mu>0.8''\, {\rm yr^{-1}}$), are so
restrictive that only a small ($1\%$) fraction of the halo white dwarf
population can be found in the selection zone.

\subsection{The thick disk contribution}

As  already  mentioned, the  thick  disk  is  characterized by  higher
velocity distributions  and a  larger scale height  than those  of the
thin disk.  Several  studies on halo white dwarfs  have considered the
thick  disk   population  as   a  possible  source   of  contamination
(Oppenheimer et al. 2001; Reid et al. 2001; Torres et al. 2002), but a
comprehensive theoretical  study remains  to be done.   Accordingly we
evaluate in  this section the  joint contribution of thick  disk white
dwarfs and red  dwarfs to the microlensing optical  depth.  We do that
for both  the MACHO and  EROS experiments in  the same way as  for the
halo  simulations presented  in the  previous section.   The  model of
spectral evolution of white dwarfs adopted for this study is our model
B, which  we consider to be  the most realistic  one.  Before starting
the discussion  of our  results, we would  like to emphasize  that the
calculation of the microlensing optical depth involves the addition of
individual  contributions,  which  are  proportional to  the  Einstein
crossing  time  corrected  by  the  efficiency function  ---  see  Eq.
(3). The  efficiency function  in turn depends  on the  crossing time,
which is  directly proportional to  the Einstein radius  and inversely
proportional  to the velocity  perpendicular to  the observer.   For a
thick  disk object  the average  distance is  smaller than  that  of a
typical  halo object. Thus  there are  two competing  effects, smaller
distances  clearly  imply  smaller  individual  contributions  to  the
optical depth.   However, thick disk stars have  also lower velocities
than  those  of the  halo,  thus  implying  more important  individual
contributions.   The   precise  balance  between   these  two  effects
determines the final contribution.

\begin{figure*}
\vspace{11.5cm}
\includegraphics{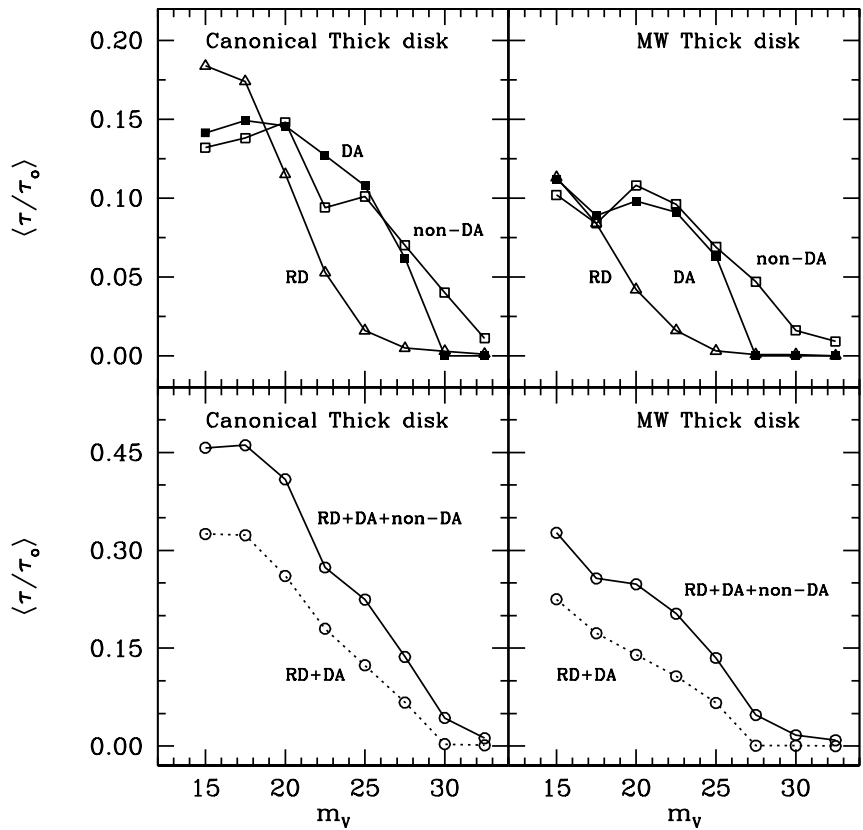}
\caption{Microlensing optical  depth towards the LMC as  a function of
         the limiting  magnitude for the thick  disk population. Solid
         and  open squares  represent the  DA and  non-DA  white dwarf
         populations,  respectively.  Red  dwarfs  are represented  by
         open triangles, while the  entire population is shown by open
         circles.}
\end{figure*}

\begin{table*}
\centering
\begin{tabular}{lrrrrrrrrrrrrr}
\hline 
\hline 
\multicolumn{1}{c}
{\ } 
& \multicolumn{4}{c}{Canonical thick disk} 
& \multicolumn{4}{c}{Metal-weak thick disk}\\
\hline 
Magnitude                                          & 17.5    & 22.5    & 27.5    & 30.0    & 17.5    & 22.5    & 27.5    & 30.0    \\
\cline{2-5}
\cline{6-9}
$\langle N_{\rm WD}\rangle$                        & $0\pm1$ & $0\pm1$ & $0\pm1$ & $0\pm1$ & $0\pm1$ & $0\pm1$ & $0\pm1$ & $0\pm1$ \\
$\langle N_{\rm RD}\rangle$                        & $0\pm1$ & $0\pm1$ & $0\pm1$ & $0\pm1$ & $0\pm1$ & $0\pm1$ & $0\pm1$ & $0\pm1$ \\ 
$\langle m_{\rm WD}\rangle$ $(M/M_{\sun})$         & 0.616   & 0.615   & 0.588   & 0.593   & 0.617   & 0.613   & 0.650   & 0.709   \\
$\langle m_{\rm RD}\rangle$ $(M/M_{\sun})$         & 0.371   & 0.221   & 0.111   & 0.085   & 0.384   & 0.269   & 0.091   & 0.079   \\ 
$\langle \eta  \rangle$                            & 0.348   & 0.348   & 0.443   & 0.999   & 0.431   & 0.452   & 0.947   & 0.999   \\  
$\langle\mu\rangle\;(''\,{\rm yr}^{-1})$           & 0.005   & 0.004   & 0.003   & 0.002   & 0.009   & 0.008   & 0.007   & 0.001   \\  
$\langle d\rangle$ (kpc)                           & 3.42    & 4.07    & 5.11    & 5.17    & 1.75    & 2.04    & 2.29    & 1.39    \\ 
$\langle V_{\rm  tan}\rangle\;({\rm km\,s}^{-1})$  & 82      & 84      & 78      & 82      & 79      & 84      & 77      & 69      \\ 
$\langle\hat{t}_{\rm E}\rangle $ (d)               & 156     & 173     & 204     & 186     & 125     & 122     & 136     & 121     \\
$\langle\tau/\tau_0 \rangle$                       & 0.462   & 0.274   & 0.137   & 0.012   & 0.257   & 0.203   & 0.050   & 0.020   \\ 
\hline
\hline
\end{tabular}
\caption{Summary of the results obtained for the thick disk population
         of microlenses towards the  LMC for the MACHO experiment with
         a thick disk age of 12~Gyr and several magnitude cuts.}
\end{table*}

The results of  these simulations are shown in Fig. 4  and Table 3. In
the top  panels of Fig.  4  we show the contribution  of the different
populations  under study  to the  optical depth  derived by  the MACHO
experiment for both  the canonical thick disk ---  left panels --- and
the  metal-weak  thick  disk  of  Carollo et  al.   (2009)  ---  right
panels. It is interesting to realize  that in both cases DA and non-DA
white dwarfs contribute by roughly the same amount.  Additionally, the
contribution  of  red  dwarfs  quickly decreases  and  becomes  almost
negligible for  realistic magnitude cuts,  while that of  white dwarfs
decreases only  slightly.  Consequently, for  realistic magnitude cuts
--- say $m_v > 20^{\rm mag}$ --- the contributions to the microlensing
optical  depth  of both  DA  and non-DA  white  dwarfs  are much  more
significant than that  of red dwarfs. In the bottom  panels of Fig.  4
we show the contribution of  the entire population to the microlensing
optical depth for both thick disk  models.  We obtain that in the case
of the MACHO experiment and  for a typical magnitude cut of $22.5^{\rm
mag}$ the contribution of the  populations of the canonical thick disk
white dwarfs  and red dwarfs to the  optical depth can be  as large as
$30\%$, which is slightly more  than that of the metal-weak thick disk
model, which is on the order  of $20\%$.  At first glance, this result
may seem to be in contrast with other recent estimates.  For instance,
Alcock et al.   (2000) estimated the contribution of  thick disk stars
to be $\sim 2\%$ of the  observed optical depth.  This agrees with our
model  if we  only consider  the red  dwarf population,  for  which we
obtain a contribution to the optical depth of $\sim 3\%$, a value very
similar to that  obtained by Alcock et al.   (2000).  On the contrary,
when thick disk white dwarfs  are taken into account, the contribution
of the thick disk is as large as that of the halo.

\begin{table*}
\centering
\begin{tabular}{lrrrrrrrrrrrrr}
\hline 
\hline 
\multicolumn{1}{c}
{\ } 
& \multicolumn{4}{c}{Canonical thick disk} 
& \multicolumn{4}{c}{Metal-weak thick disk}\\
\hline 
Magnitude                                          & 17.5    & 22.5    & 27.5    & 30.0    & 17.5    & 22.5    & 27.5    & 30.0    \\
\cline{2-5} 
\cline{6-9}
$\langle N_{\rm WD}\rangle$                        & $0\pm1$ & $0\pm1$ & $0\pm1$ & $0\pm1$ & $0\pm1$ & $0\pm1$ & $0\pm1$ & $0\pm1$ \\
$\langle N_{\rm RD}\rangle$                        & $0\pm1$ & $0\pm1$ & $0\pm1$ & $0\pm1$ & $0\pm1$ & $0\pm1$ & $0\pm1$ & $0\pm1$ \\ 
$\langle m_{\rm WD} \rangle$ $(M/M_{\sun})$        & 0.629   & 0.636   & 0.651   & 0.746   & 0.584   & 0.595   & 0.619   & 0.604   \\
$\langle m _{\rm RD}\rangle$ $(M/M_{\sun})$        & 0.200   & 0.221   & 0.106   & 0.080   & 0.325   & 0.177   & 0.085   & 0.076   \\ 
$\langle\eta\rangle$                               & 0.297   & 0.425   & 0.667   & 0.999   & 0.394   & 0.438   & 0.667   & 0.999   \\  
 $\langle\mu\rangle\;(''\,{\rm yr}^{-1})$          & 0.005   & 0.004   & 0.004   & 0.006   & 0.010   & 0.009   & 0.007   & 0.008   \\  
$\langle d\rangle$ (kpc)                           & 3.37    & 4.08    & 4.41    & 3.73    & 1.68    & 1.95    & 2.55    & 2.09    \\ 
$\langle V_{\rm  tan}\rangle\;({\rm km\,s}^{-1})$  & 87      & 88      & 94      & 98      & 83      & 85      & 86      & 77      \\ 
$\langle\hat{t}_{\rm E}\rangle $ (d)               & 137     & 141     & 149     & 140     & 110     & 112     & 141     & 110     \\
$\langle\tau/\tau_0\rangle$                        & 1.360   & 1.214   & 0.529   & 0.083   & 1.224   & 0.546   & 0.308   & 0.015   \\ 
\hline
\hline
\end{tabular}
\caption{Same as table 3 for the EROS experiment.}
\end{table*}

A more detailed analysis of the  thick disk population can be done and
reveals  that in the  case of  the canonical  thick disk  the possible
microlensing   events   have  an   Einstein   crossing  time   $t_{\rm
E}\approx\,170$ days  for a magnitude  cut of $22.5^{\rm  mag}$, while
for  the  case of  the  metal-weak  thick  disk the  average  Einstein
crossing time  amounts to  $t_{\rm E}\approx\,120$ days.   Both values
are considerably  higher than that of the  halo population.  Moreover,
as can  be seen in  Table 3, the  mean average tangential  velocity is
$\sim 80\;{\rm km\;s^{-1}}$ for both models --- which is what we would
expect for a  thick disk population, but the  mean average distance of
the lenses  is $\sim 4$ kpc  --- which is comparable  to that obtained
for the  halo population.  This can  be easily understood  in terms of
the selection criteria we use to  decide when a star can be considered
responsible of a microlensing  event.  In particular, we only consider
as reliable microlensing events those in which the lens is dimmer than
a  certain magnitude  cut  and,  given that  the  thick population  is
intrinsically brighter than the  halo population, we only select those
thick disk lenses  which are far enough away.  In any  case, as can be
seen in Table 3, our  simulations show that the thick disk populations
can produce  at most  one microlensing event.   We emphasize  that the
results  obtained  using the  canonical  thick  disk  model appear  to
provide an upper limit for  the contribution to the total microlensing
optical depth --- see Fig. 4.

We have  also estimated the contribution  to the optical  depth of the
thick disk populations in the case of the EROS experiment. The results
are  shown in Table  4. For  a realistic  magnitude cut  of $22.5^{\rm
mag}$,  $1\pm1$   microlensing  event  is  expected   at  a  $1\sigma$
confidence level.   The confirmation of this  microlensing event would
increase the value  of the optical depth measured by  the EROS team by
$\sim 40\%$.   However, we point  out that given the  poor statistics,
the number  of microlensing events obtained in  our simulations agrees
reasonably  well with  the observations  of the  EROS team,  who found
none.

\subsection{The event rate distribution}

\begin{figure*}
\vspace{12.5cm}
\includegraphics{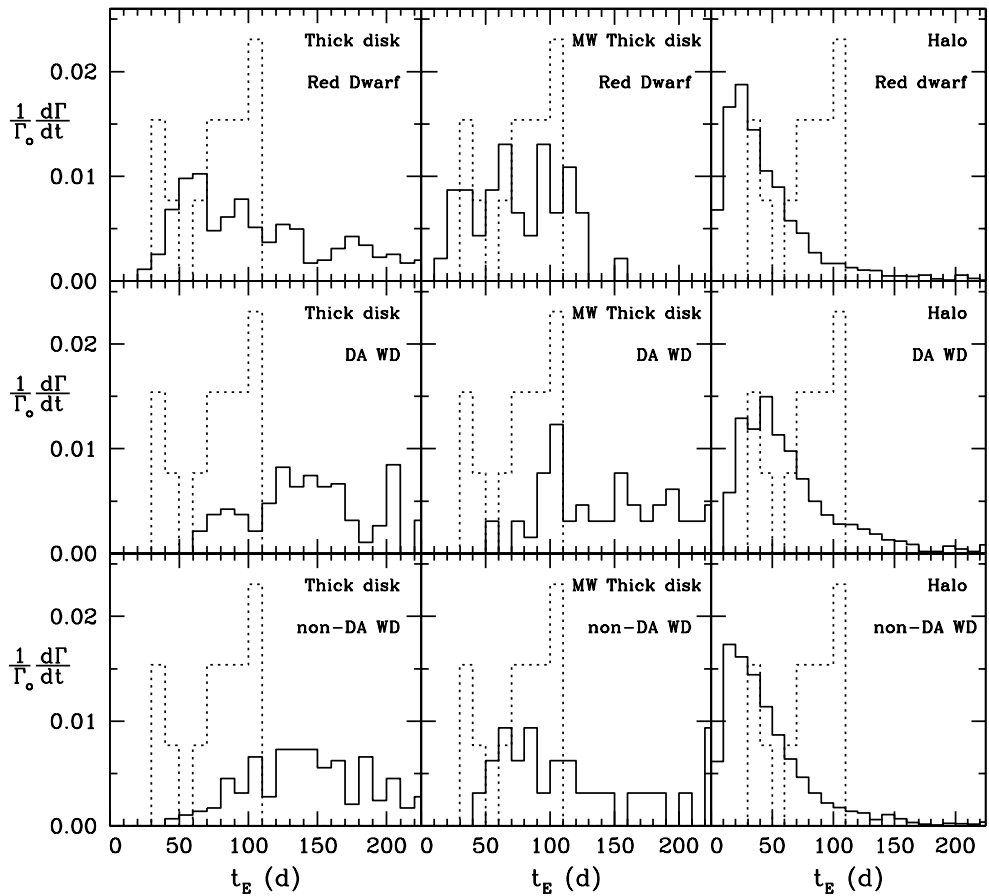}
\caption{Normalized distributions of the  microlensing event rate as a
         function of  the event duration for  the different population
         under  study   (solid  lines)  and  also   for  the  observed
         microlensing events  of the MACHO  experiment (dashed lines).
         All the distributions have  been computed for a magnitude cut
         of 22.5$^{\rm mag}$.}
\end{figure*}

Besides  the optical  depth and  the Einstein  crossing time,  a third
quantity was proposed by Paczy\'nski  (1986) as relevant for the study
of the  microlensing experiments.   This quantity is  the microlensing
event rate $\Gamma$, which provides the rate at which the lenses enter
the microlensing tube. While the  optical depth does not depend on the
mass  function,   the  event   rate  does,  and   consequently  useful
information  about  the  different  populations  responsible  for  the
micronlensing envents can be obtained by studying it.

In Fig. 5 we display  with solid lines the normalized distributions of
the microlensing  event rate as a  function of the  event duration for
the different  populations of the canonical thick  disk (left panels),
the  metal-weak  thick  disk  (central  panels) and  the  halo  (right
panels). We  also show the results  obtained by the MACHO  team with a
dashed line.  The top  panels show the  distribution obtained  for the
population of red dwarfs, while  the middle and the bottom panels show
the distributions obtained for the  populations of DA and non-DA white
dwarfs respectively. To produce  these distributions we have adopted a
magnitude  cut  of  $22.5^{\rm  mag}$,  which  can  be  considered  as
representative  of   the  current  experiments.    Although  for  this
magnitude cut  the contribution to  the microlensing optical  depth is
approximately the same  for the canonical thick disk  (27.4\%) and the
halo  populations (24.9\%)  and  somewhat smaller  for the  metal-weak
thick disk  (20.3\%), the  event rate distributions  are significantly
different.   We  obtain for  instance  for  the  canonical thick  disk
population in all the cases  very extended distributions with no clear
maxima (except  for the  population of red  dwarfs, which peaks  at 70
days), while for the metal-weak  thick disk population of red dwarfs a
deficit of stars with long Einstein crossing times is found.  Finally,
for  the  halo  populations   the  theoretical  distributions  of  the
microlensing event  rate have  clear maxima and  moderate dispersions.
Still, all  the distributions are only marginally  consistent with the
observational distribution, which is characterized by event timescales
in the range of 35 to  110 days.  These results show the difficulty of
explaining the MACHO microlensing events with a single population.

A more  quantitative assesment of  the compatibility of  the different
simulated populations  with the observational  data can be  done using
the $Z^2$  compatibility test. The $Z^2$ statistical  test (Lucy 2000)
is specifically  designed to  deal with meagre  data sets and  is thus
especially   well  suited  for   our  case.    The  results   of  this
compatibility  test  are  presented  in  table  5  for  the  different
populations under  study and for  different magnitude cuts. As  can be
seen, the probability that for our reference magnitude cut ($22.5^{\rm
mag}$) the  distributions of microlensing timescales  of the canonical
thick disk populations are  compatible with the observational data are
$\sim 0.8$, while those of the metal-weak thick disk are less probable
($\sim  0.7$) and  those of  the halo  populations  are signifincantly
smaller,  $\sim  0.6$.  In  particular  it is  to  be  noted that  the
population of  red dwarfs  and DA white  dwarfs of a  metal-weak thick
disk is  practically incompatible with  the observational data  when a
magnitude cut of $30^{\rm  mag}$ is adopted.  As previously mentioned,
the thick disk populations  present wider timescale distributions than
the halo (see  Fig.  5) and, given that  the observational results are
to some extent  spread over a wide range of  timescales, the result is
that the  thick disk populations match  the observational distribution
better.  Finally, it  is interesting to realize that  as the magnitude
cut  becomes larger, the  compatibility of  the thick  disk population
with the  observational data  substantially decreases, while  those of
the halo  white dwarf  populations remain almost  constant. That  is a
clear consequence of  the different nature of the  thick disk and halo
white  dwarf populations:  while the  population of  thick  disk white
dwarfs  is constituted  by  bright objects  at  sufficiently far  away
distances, the  halo white dwarf  population is made  of intrinsically
dim and distant objects.

\begin{table}[t]
\smallskip
\centering
\centering {$Z^2$ COMPATIBILITY TEST}\\
\vspace{0.25cm}
\begin{tabular}{lcccc}
\hline
\hline
\multicolumn{1}{c}{\ } &
\multicolumn{4}{c}{Canonical thick disk} \\
\cline{2-5}
Magnitude              & 17.5 & 22.5 & 27.5 & 30.0 \\ 
\cline{2-5} 
Red dwarfs             & 0.87 & 0.86 & 0.64 & 0.39 \\
DA white dwarfs        & 0.84 & 0.82 & 0.68 & 0.54 \\
non-DA white dwarfs    & 0.84 & 0.86 & 0.82 & 0.67 \\
\hline
\multicolumn{1}{c}{\ } &
\multicolumn{4}{c}{Metal-weak thick disk} \\
\cline{2-5}
Magnitude              & 17.5 & 22.5 & 27.5 & 30.0 \\ 
\cline{2-5} 
Red dwarfs             & 0.84 & 0.73 & 0.43 & 0.08 \\
DA white dwarfs        & 0.76 & 0.74 & 0.74 & 0.01 \\
non-DA white dwarfs    & 0.70 & 0.73 & 0.70 & 0.41 \\
\hline
\multicolumn{1}{c}{\ } &
\multicolumn{4}{c}{Halo} \\
\cline{2-5}
Magnitude              & 17.5 & 22.5 & 27.5 & 30.0 \\ 
\cline{2-5} 
Red dwarfs             & 0.54 & 0.57 & 0.43 & 0.37 \\
DA white dwarfs        & 0.56 & 0.62 & 0.71 & 0.61 \\
non-DA white dwarfs    & 0.52 & 0.56 & 0.56 & 0.60 \\
\hline
\hline
\end{tabular}
\caption{$Z^2$ compatibility test  between the Einstein crossing times
         obtained by the  MACHO team and the simulated  thick disk and
         halo populations for different magnitude cuts.}
\end{table}

\subsection{The joint contribution of the thick disk and halo populations}

We  have also  computed  the joint  contribution  to the  microlensing
optical depth of the thick disk and halo populations of red dwarfs and
white dwarfs.  A summary of our  results is displayed in Table 6 where
the same quantities are displayed as  those in Table 1 for the case of
the  halo population  and  Table 3  for  the case  of  the thick  disk
simulation.   We present the  outcome for  both the  cases in  which a
canonical thick disk (left section) and a metal-weak thick disk (right
section)  are adopted.   Moreover,  we  only show  in  this table  the
results obtained  for model B (which  we remind is  the most realistic
model for  the white dwarf population)  when the results  of the MACHO
towards the  LMC are  simulated.  The main  effect is that  the entire
population under  study can  explain about half  of the  optical depth
obtained by  the MACHO experiment,  irrespective of the  adopted thick
disk model. This value practically  doubles the one obtained when only
the halo  population was  considered. Clearly, these  results indicate
that  the  thick disk  population  must be  taken  into  account as  a
potential source  of contamination in the current  experiments.  For a
realistic  magnitude  cut of  $22.5^{\rm  mag}$,  a  maximum of  three
microlensing  events is  expected at  the $1\sigma$  confidence level,
which  is far below  the $\sim  11$ microlensing  events of  the MACHO
experiment.  Our results show that in  the case in which a white dwarf
is  responsible of  the microlensing  event, the  average mass  of the
lenses is approximately $0.6\,M_{\sun}$,  whereas in the case in which
the  one responsible for  the microlensing  event is  a red  dwarf the
average mass  is $\sim  0.2\, M_{\sun}$.  Both  values are  typical of
their respective populations.  Moreover,  in the case of white dwarfs,
half of the microlensing events are due to non-DA white dwarfs.

\begin{table*}
\centering
\begin{tabular}{lrrrrrrrrrrrr}
\hline 
\hline 
\multicolumn{1}{c}{\ } & 
\multicolumn{4}{c}{Canonical thick disk+halo} &
\multicolumn{4}{c}{Metal-weak thick disk+halo}\\
\hline 
Magnitude                                         & 17.5    & 22.5    & 27.5    & 30.0    & 17.5    & 22.5    & 27.5    & 30.0    \\ 
\cline{2-5}
\cline{6-9}
$\langle N_{\rm WD}\rangle$                       & $0\pm2$ & $0\pm2$ & $0\pm2$ & $0\pm2$ & $0\pm2$ & $0\pm2$ & $0\pm2$ & $0\pm2$ \\
$\langle N_{\rm RD}\rangle$                       & $1\pm2$ & $1\pm2$ & $0\pm2$ & $0\pm2$ & $1\pm2$ & $1\pm2$ & $0\pm2$ & $0\pm2$ \\
$\langle m_{\rm WD} \rangle$ $(M_{\sun})$         & 0.613   & 0.612   & 0.596   & 0.607   & 0.614   & 0.610   & 0.644   & 0.697   \\
$\langle m _{\rm RD}\rangle$ $(M_{\sun})$         & 0.360   & 0.222   & 0.112   & 0.084   & 0.368   & 0.258   & 0.098   & 0.079   \\
$\langle\eta\rangle$                              & 0.446   & 0.450   & 0.546   & 0.998   & 0.528   & 0.548   & 0.944   & 0.998   \\
$\langle\mu\rangle$ $(''\,{\rm  yr}^{-1})$        & 0.008   & 0.006   & 0.005   & 0.005   & 0.120   & 0.010   & 0.009   & 0.004   \\
$\langle d\rangle$ (kpc)                          & 3.20    & 3.89    & 4.78    & 4.85    & 1.89    & 2.35    & 2.59    & 2.01    \\
$\langle V_{\rm tan}\rangle$ $({\rm km\,s}^{-1})$ & 115     & 118     & 114     & 120     & 119     & 124     & 121     & 117     \\
$\langle\hat{t}_{\rm E}\rangle$ (d)               & 131     & 146     & 171     & 159     & 103     & 103     & 114     & 106     \\ 
$\langle\tau/\tau_0\rangle$                       & 0.738   & 0.523   & 0.316   & 0.093   & 0.533   & 0.460   & 0.238   & 0.105   \\ 
\hline 
\hline
\end{tabular}
\caption{Summary of the  results obtained for the thick  disk and halo
         populations  of microlenses  towards  the LMC  for the  MACHO
         experiment.}
\end{table*}

\subsection{Halo dark matter}

We have also computed the  contribution of the halo populations to the
baryonic  dark matter  density of  the Galaxy.   The fraction  of dark
matter in the  form of MACHOs, $f$, can be  directly obtained from the
microlensing  optical   depth  towards  the  LMC.    Assuming  a  halo
isothermal sphere we have $\tau_{\rm LMC}=5.1\times 10^{-7}f$. Thus we
obtain from our  simulations $f=0.07$ in the case  of model A, whereas
for model B we derive $f=0.06$.  These values can be compared with our
previous  results of  $f=0.05$ (Torres  et al.   2008).  Thus  when we
include  the  population of  non-DA  white  dwarfs  we find  a  modest
increase, which  is still not enough  to account for the  bulk of halo
dark matter.


\section{Conclusions}

We have  analyzed the contribution  to the microlensing  optical depth
towards  the LMC  of the  halo population  of white  dwarfs  with both
hydrogen-rich and hydrogen-deficient  atmospheres.  We have used three
models to describe the atmospheric  evolution of white dwarfs.  In the
first  of  these   models  we  have  assumed  a   canonical  ratio  of
hydrogen-rich  white  dwarfs, to  80\%  independent  of the  effective
temperature.  In  our second model, which  we consider to  be the most
realistic  one,  we have  adopted  a  fraction  of white  dwarfs  with
helium-rich atmospheres which depend on the effective temperature.  In
these two models the cooling sequences of Salaris et al. (2000) for DA
white dwarfs and those of Benvenuto \& Althaus (1997) for non-DA white
dwarfs were  used. In  the third set  of calculations the  fraction of
non-DA  white   dwarfs  was  assumed   to  depend  on   the  effective
temperature,  but the  cooling tracks  of Bergeron  et al.  (1995) for
non-DA  white  dwarfs were  adopted.   We  have  found that  when  the
contribution  of hydrogen-deficient  white dwarfs  is  considered, the
theoretical optical depth towards the  LMC for both the MACHO and EROS
experiments is substantially increased  by nearly $34\%$, with respect
to  previous calculations. Nevertheless,  we have  also found  that no
more than  one third  of the microlensing  optical depth found  by the
MACHO team can be explained by  the halo population of white dwarfs at
the  $1\sigma$   confidence  level,  and  that  no   more  than  three
microlensing events could be expected  at the same confidence level in
reasonable agreement with the results of the EROS experiment.

We have also studied the role  played by the thick disk populations of
white dwarfs and red dwarfs, thus extending our previous calculations.
For this purpose we have used two thick disk models.  The first one is
a canonical thick disk model,  while the second one corresponds to the
most recent model of Carollo et al. (2009), which is based in the data
of the Sloan Digital Sky Survey Data Release 7.  We have obtained that
for both thick  disk models, the contribution of  these populations to
the  microlensing optical  depth is  comparable  to that  of the  halo
populations,  which is somewhat  larger for  the canonical  thick disk
model,  which  provides  an  upper  limit to  this  contribution.   In
particular we have found that the thick disk contribution is dominated
by the  white dwarf population in  both cases, as  the contribution of
thick  disk red  dwarfs  is only  half  of that  of  halo red  dwarfs.
Besides, we have also found that the average distance of the simulated
lenses is  very similar for the  thick and halo  populations, $\sim 3$
kpc.  This unexpected  result can be easily explained  in terms of the
selection  criteria used  to  decide  when a  star  can be  considered
responsible  of  a microlensing  event.   Since  we  only consider  as
reliable microlensing events those in  which the lens is dimmer than a
certain magnitude cut, intrinsically  bright lenses must be located at
larger  distances.   Consequently,   since  the  thick  population  is
intrinsically brighter than the  halo population, we only select those
thick disk lenses which are far enough away, at distances very similar
to those of the halo  population, which are naturally located at large
distances.  We have found as  well that although both populations have
similar average distances and  thick disk objects have smaller average
velocities,  their  event  timescales  are  nearly  three  times  more
extended than those of the halo population.  We have also assessed the
compatibility   of   our  simulated   populations   with  the   scarce
observational  data.  We  have found  that the  thick  disk population
agrees  better  with the  MACHO  observational  distribution of  event
timescales than the halo population.

Finally,  we  found  that  when  both  the halo  and  the  thick  disk
populations are considered,  nearly half of the measured  value of the
microlensing optical  depth towards  the LMC can  be explained  at the
95\%  confidence   level  by  our   simulated  halo  and   thick  disk
populations.  According to our  simulations, the fraction of halo dark
matter  that can  be  expected from  MACHOs  increases moderately  (to
$f=0.06$)   with   respect    to   our   previous   simulations   when
hydrogen-deficient white dwarfs are taken into account.


\begin{acknowledgements}
Part   of    this   work   was    supported   by   the    MEC   grants
AYA08--04211--C02--01 and AYA08--1839/ESP, by the European Union FEDER 
funds and by the AGAUR.
\end{acknowledgements}



\begin{thebibliography}{1}
\bibitem{AL96}  Adams, F.C., \& Laughlin, G., 1996, ApJ, 468, 686
\bibitem{AEA97} Alcock, C., Allsman, R.A., Alves, D.R., et  al., 1997,
                \apj, 486, 69
\bibitem{AEA00} Alcock, C., Allsman,  R.A., Alves, D.R., et al., 2000,
                \apj, 542, 281
\bibitem{AEA07} Althaus,  L.G., Garc\'\i  a--Berro, E., Isern,  J., et
                al., 2007, \aap, 465, 249
\bibitem{BEA98} Baraffe,  I., Chabrier, G., Allard, F.,  et al., 1998,
                \aap, 337, 403
\bibitem{BEA05} Becker, A.C.  Rest, A.   Stubbs, C., et al., 2005, IAU
                Symposium, 225, 357
\bibitem{BA97}  Benvenuto,  O.G., \& Althaus, L.G., 1997, \mnras, 288, 
                1004
\bibitem{BL02}  Bergeron, P., \& Leggett, S. K., \apj 580, 1070
\bibitem{BLR01} Bergeron,  P., Leggett, S.  K., \& Ruiz, M.  T., 2001,
                \apjs, 133, 413
\bibitem{BWB95} Bergeron, P., Wesemael, F.,  \&  Beauchamp, A.,  1995,
                \pasp, 107, 1047
\bibitem{B87}   Binney,  J.,  \&  Tremaine,  H., 1987,  {\sl  Galactic
		Dynamics} (Princeton: Princeton Univ.  Press)
\bibitem{CEA07} Camacho,  J.,  Torres, S.,  Isern, J.,  et  al., 2007,
                \aap, 471,151
\bibitem{C97}   Canal,  R.,  Isern,  J.,  \& Ruiz--Lapuente, P., 1997, 
                \apj, 488, L35
\bibitem{CEA09} Carollo,  D., Beers,  T.C., Chiba,  M., et  al., 2009,
                \apj, in press, {\tt arXiv:0909.3019}
\bibitem{Cea96} Chabrier, G., Segretain, L., \& M\'era, D., 1996, ApJ, 
                468, 21
\bibitem{CH95}  Charlot, S., \& Silk, J., 1995, \apj, 445, 124
\bibitem{DB98}  Dehnen, W.  \& Binney, J., 1998, \mnras, 298, 387
\bibitem{FEA01} Flynn,  C.,  Sommer--Larsen,  J., Fuchs,  B., et  al.,
                2001, \mnras, 322, 553
\bibitem{GEA99} Garc\'\i a--Berro, E.,  Torres, S., Isern, J., et al.,
                1999, \mnras, 302, 173
\bibitem{GEA04} Garc\'\i a--Berro, E.,  Torres, S., Isern, J., et al.,
                2004, \aap, 418, 53
\bibitem{GGA04} Gates,  E.,  Gyuk, G.,  Harris, H.  C., et  al., 2004,
                \apj, 612, L129-L132
\bibitem{GWJ95} Gilmore, G., Wyse,  R.F.G.  \& Jones, J.B., 1995, \aj,
                109, 1095
\bibitem{GEA02} Goldman,  B., Afonso,  A., Alard,  Ch., et  al., 2002,
		\aap, 389, 69
\bibitem{G91}   Griest, K., 1991, \apj, 366, 412
\bibitem{GG099} Gyuk, G., \& Gates, E., 1999, \mnras, 304, 281
\bibitem{GDG00} Gyuk, G.,  Dalal, N., \& Griest, K.,  2000, \apj, 535,
                90
\bibitem{H98}   Hansen, B.M.S., 1998, \nat, 394, 860
\bibitem{HL03}  Hansen, B.M.S., \& Liebert, J., 2003, \araa, 41, 465
\bibitem{HEA06} Harris,  H.C., Munn,  J.A., Kilic,  M., et  al., 2006,
                \aj, 131, 571
\bibitem{I89a}  Iben, I., \& Laughlin, G., 1989, \apj, 341, 312
\bibitem{I98}   Isern, J., Garc\'\i a--Berro, E., Hernanz, M., et al.,
                1998, \apj, 503, 239
\bibitem{J90}   James, F., 1990, Comput.  Phys.  Commun., 60, 329
\bibitem{K06}   Kallivayalil, N., van der Marel, R.P., Alcock,  C., et
                al., 2006, \apj, 638, 772
\bibitem{KEA06} Kilic,  M., Munn,  J.A., Harris,  et al.,  2006, \apj,
                131, 582
\bibitem{LEA01} Lasserre, T.  Afonso.  C., Albert, J.N., et al., 2001,
		\aap, 355, L39
\bibitem{L89}   Liebert,  J., 1989, in {\sl ``White Dwarfs''}, Ed.: G. 
                Wegner (Berlin: Springer), 15
\bibitem{L00}   Lucy, L., 2000, \mnras, 318, 92
\bibitem{MR02}  Mollerach, S., \& Roulet, E., 2002, {\sl Gravitational
                Lensing    and    Microlensing}   (Singapore:    World
                Scientific)
\bibitem{MEA99} Muraki, Y., Sumi, T.,  Abe, F., et al., 1999, Progress
                of Theoretical Physics Supplement, 133, 233
\bibitem{NFW}   Navarro, J.F.,  Frenck, C.S., \& White,  S.D.M., 1997,
                \apj, 490, 493
\bibitem{OEA01} Oppenheimer, B.R., Hambly,  N.C., Digby, A.P., et al.,
                Science, 292, 698
\bibitem{P86}   Paczy\'nski, B., 1986, \apj, 304, 1
\bibitem{R05}   Reid, I., N., 2005, \araa, 43, 247
\bibitem{RSH01} Reid, N.I., Sahu, K.C., \& Hawley,  S.L.,  2001,  \apj,
		559, 942
\bibitem{SEA00} Salaris,   M.,  Garc\'{\i}a--Berro, E.,  Hernanz,  M.,
		et al., 2000, \apj, 544, 1036
\bibitem{S98}   Scalo, J., 1998, in  {\sl ``The  Stellar  Initial Mass
		Function''},  Eds.:  G.  Gilmore  \&  D.  Howell  (San
		Francisco: PASP Conference Series), Vol.  142, 201
\bibitem{SKW04} Schneider,   P., Kochanek,  C.S.,  \& Wambsganss,  J.,
                2004,  in {\sl  ``Saas-Fee  lectures on  Gravitational
                Lensing''}
\bibitem{TEA07} Tisserand,  P., Le  Guillou, L.,  Afonso, C.,  et al.,
                2007, \aap, 469, 387
\bibitem{TEA98} Torres,  S., Garc\'\i  a--Berro,  E.,  \&  Isern,  J.,
		1998, \apj, 508, L71
\bibitem{TEA02} Torres,  S.,  Garc\'\i a--Berro,  E., Burkert,  A., et
                al., 2002, \mnras, 336, 971
\bibitem{TEA08} Torres,  S.,  Camacho, J.,  Isern, J.,  et  al., 2008,
                \aap, 486, 427
\bibitem{UEA94} Udalski, A., Szymanski, M., Kaluzny, J., et al., 1994,
                Acta Astron., 44, 1
\bibitem{VEA07} Vidrih, S., Bramich, D.M., Hewett, P.C., et al., 2007,
                \mnras, 382, 515
\end{thebibliography}
\end{document}